\documentstyle{article}

\def\pa{\parallel}
\def\pe{\perp}

 \let\b=\beta   \let\d=\delta  \let\e=\varepsilon
 \let\g=\gamma \let\h=\eta      \let\l=\lambda
\let\m=\mu            
\let\r=\varrho  \let\s=\sigma \let\t=\tau   
  \let\z=\zeta
\let\D=\Delta   \let\G=\Gamma  \let\L=\Lambda 
\let\O=\Omega
\newcommand{\ie}{\hbox{\it i.e.\ }}
\title{\bf Continuum Field Description of Driven Lattice Gases}

\author{F. de los Santos and P.L. Garrido}
\date{Instituto Carlos I de F\'\i sica Te\'orica y Computacional \\
Universidad de Granada, E-18071 Granada, Spain}

\begin{document}
\maketitle
\begin{abstract}
We review the critical behaviour of the Driven Lattice Gas (DLG) model. 
As a result, we obtain a novel Langevin 
equation for the DLG which depends on the microscopic
transition probabilities. We then 
show how this dependence affects the critical behaviour of the
the DLG placing the finite and the infinite driving field cases
into different universality classes.
Two other well known anisotropic, conservative, non-equilibrium models,
the two-temperature model and 
the randomly driven model (RDLG)
are also studied. It is shown that the RDLG with infinite averaged
field and the two-temperature model with $T_\pa =\infty$ fall
in the same universality class as the infinitely driven DLG. A Langevin
equation for the two-layer DLG is also presented. 

\end{abstract}
\section{Introduction}

Equilibrium statistical mechanics has succeeded in predicting the
collective behaviour of a system in thermal equilibrium 
given the laws governing its microscopic behaviour. Unfortunately, 
most natural phenomena belong to the field of {\em non-equilibrium}
statistical mechanics, a subject far less well understood and still
in its developing stage. Being
overwhelmed by the enormous complexity of non-equilibrium systems, it
is expedient to focus on those which settle into non-equilibrium
{\em steady states} and to trace its behaviour to specific model
ingredients.
Then, results pertaining to collective behaviour can be obtained 
by several  methods. For instance, Monte Carlo simulations
in which
a great deal of our understanding of
the large scale properties of a system is based
and also serve to test predictions based on other approaches. Mean field
theories, the usual starting point of the analytic route, can provide
us with some insight into the phase diagram, but they fail to describe
the true collective behaviour properly, specially in systems with
local interactions in low dimensionalities. 
Further analytical developments are seriously hampered
by, among other things, the (commonly) discrete structure of the
system, so to describe 
the physics at the macroscopic level a new avenue is required.
To accomplish this goal one can benefit from the {\em mesoscopic}
approach in terms of Langevin equations, which concentrates on the long-time,
large-distance properties and tries to eliminate the lattice altogether, by making the order parameter into a continuous field. This mesoscopic picture
can be derived, at least in principle, starting from the 
microscopic master equation and implementing a coarse graining
procedure, but in practice this route proves to be an insurmountable task for most
systems. This predicament is usually overcome by postulating 
phenomenological equations based on the choice of an order parameter
and the underlying system symmetries. When equilibrium critical phenomena 
are considered, one can then appeal to the framework of the renormalization 
group, and {\em universality} then 
appears in the sense that the results are independent
of the microscopic details, in particular the microscopic dynamic
rules \cite{grins}. 
But non-equilibrium critical phenomena is still a challenging
matter which displays striking features \cite{MD}.
More specifically,
in contrast to equilibrium, in non-equilibrium 
situations the transition rates are not a simply matter of 
convenience.
The observable critical behaviour can depend on some details of the 
microscopic dynamics, a fact which is often underestimated in the literature. 
An analysis concluding about the relevant features that characterize the universal
properties of a non-equilibrium system at criticallity is still lacking.

This paper is devoted to the {\em driven lattice gas} 
(DLG henceforth) model and other related models. First devised by Katz, Lebowitz and Spohn \cite{KLS},
the DLG is a kinetic lattice gas of interacting particles
subject to an external uniform field. A more detailed description
of the DLG will be presented in the second section of this paper, but for
the time being let us remark 
that various generalizations of the DLG have been argued to be relevant 
to the understanding of 
a wealth of varietes of natural phenomena (see \cite{MD} for a review). 
On the theoretical front, the DLG is one of the simplest 
non-equilibrium models that is hoped to serve as a paradigm for the behaviour 
of those systems which do not posses a thermodynamic equilibrium state.
It also seems to capture the essence of strongly anisotropic systems. 
But before considering the DLG as a kind of non-equilibrium 
``metamodel" it is necessary to clarify some 
controversial issues. A longstanding glaring discrepancy exists on the value of the order 
parameter critical exponent $\b$. From Monte Carlo data for an infinite field 
(no jumps against the drive are allowed) a value close to 
1/3 is found \cite{vym} but, however, $\b=1/2$ is obtained from a mesoscopic 
Langevin approach \cite{lyc,jys}. Several efforts aiming at solving this problem 
\cite{MAGA,LW} have come to grief and the question remains under discussion. Besides this 
disagreement, the failure of the Cahn-Hilliard approach to coarsening 
dynamics \cite{allz} also casts some doubts on the reliability of the standard 
continuum approach to the DLG. Open questions abound and clearly new inroads on this 
subject are needed. So, seeking a new framework for answers, 
we propose a new approach whose preliminary results were presented in
\cite{pre}.

The remainder of this paper is organized as follows:
In the first part of Section 2 we briefly summarize the basic 
ingredients of the DLG.  
We then give our derivation of a novel Fokker-Planck
equation for the DLG and its stochastically equivalent Langevin equation.
In Section 3 the 
two-temperature model, the Randomly Driven Lattice Gas and
the multilayer variant of the DLG are studied employing the 
formalism of Section 2.
Our concluding remarks are in Section 4.

\section{From Driven Lattice Gases to Driven Diffusive Systems}

Consider a set of particles confined to a box, $\lambda \subset Z^d$,
with periodic boundary conditions.
A configuration of this system is specified by giving all the 
site occupation variables, $n_{\bf x}=1,0$, 
echoing the fact that a particle may be present or not at site 
$\bf x$. Besides this hard-core constraint, the model also includes 
a nearest-neighbour (NN) interaction so, given a 
configuration $C=\{n_{\bf x}\}_{\bf x\in \lambda}$, the Hamiltonian reads
\begin{equation}
H(C)=-J\sum_{NN} n_{\bf x}n_{\bf y}.
\label{ham}
\end{equation}
The configurations $C$ evolve according to a stochastic hopping dynamics 
which conserves the number of particles, or equivalently the 
density $\r$.  
Up to the present, all we have is the familiar kinetic 
lattice gas \cite{gas} for which the following {\em master equation} 
for the time evolution of the probability distribution $P_t(C)$ applies

\begin{equation}
\partial_t P_t(C) =\sum_{C'} \Big\{ W[C' \to C]P_t(C')-W[C \to C']P_t(C)
\Big\}.
\label{me}
\end{equation}
Here, $W[C \to C']$ stands for the rate at which the system makes a 
transition from $C$ to $C'$, and $C$ and $C'$ can differ by a single 
nearest-neighbour particle-hole exchange. The choice of 
$W[C \to C']=D(\b \triangle H)$, where $D$ is any function satisfying 
$D(-x)=e^xD(x)$
and $\triangle H=H(C')-H(C)$, ensures that the stationary solution of 
(\ref{me}) is the equilibrium
one, \ie $P(C) \propto e^{-\beta H(C)}$, and $\b=1/T$ is the
inverse of the temperature of the thermal bath. 

Next, let us introduce a uniform (in both space and time) 
external drive $\bf E$ pointing along one of the principal axis of the 
lattice. We refer to it as the {\em electric} field while imagining 
that the particles behave as positive ions only in relation to it. 
The field 
biases the rates favouring jumps along its direction, suppressing 
jumps against it, and leaving unaffected those in the tranverse 
directions. For hard wall boundary conditions the only effect of the 
drive would be to add a gravitational potential energy to the 
Hamiltonian (\ref{ham}), and the resulting steady state would be an equilibrium one.
But due to the periodic boundary conditions, the drive has a 
dramatical effect on the system static properties, preventing the system 
from achieving an equilibrium stationary state. 
In this case, the electric force is 
nonconservative so it is not derivable from a global potential. 
However, the local effect of the drive could be mirrored by adding to 
the Hamiltonian (\ref{ham}) the work done by the field during the jump, so we 
choose the rates in the form

\begin{equation}
W[C \to C']=D(\b \triangle H +\b \ell E),
\end{equation}
where $\ell=(1,0,-1)$ for jumps (against, transverse to, along) $\bf E$,
and $E$ is the strength of the electric field ($E=|\bf E|)$. 
When $E=0$, one recovers the familiar equilibrium rates. 

Let us end this r\'esum\'e of the main features of the DLG (see \cite{MD,syz} for 
fairly detailed reviews) with a few  words on its collective behaviour. 
At half-filling ($\r=0.5$) and $E=0$, a second 
order phase transition at the Onsager critical 
temperature, $T_c$, is known to occur in two
dimensions. For $E \neq 0 $ and still at half-filled lattice, 
from the gleanings provided by 
Monte Carlo simulations, the DLG undergoes a second order phase transition at a 
higher critical
temperature, $T_c^{(E)}$, saturating at about 
$T_c^{(\infty)}=1.4T_c$ for $E\rightarrow\infty$. 
The main observation is that for temperatures above $T_c^{(E)}$ the particles 
are distributed homogeneously while below  
$T_c^{(E)}$ the 
DLG segregates into a particle-poor phase and a particle-rich region, the latter having 
domain walls parallel to the field. 

\subsection{The continuum limit}
In this paper we are mainly concerned with the continuum field description of
the 
DLG, so firstly we aim at making the order parameter into a continuous field.
To serve this purpose, let us mentally construct around each 
lattice point a boxi consisting of $\O^d$ 
sites. We then define a coarse-grained density $\phi_{\bf r}, {\bf r} \in \lambda$ 
through
\begin{equation}
\phi_{\bf r}={1 \over \O^d} \sum_{{\bf x} \in \O_{\bf r}} n_{\bf x}\quad ,
\end{equation}
where $\O_{\bf r}$ is the box centred at {\bf r}. 
Now, we find inspiration in the original 
occupation variables dynamics to postulate the time evolution of the new density 
variables: we perform 
exchanges between two randomly chosen nearest neighbour sites, 
$\bf r$ and $\bf r + \bf a$. 
Hence, if $C^{{\bf r}a}$ is the configuration after the exchange, 
we have
\begin{equation}
C^{{\bf r}a} =\{\phi_{\bf x} + \O^{-d}(\d_{\bf x, r+a}-\d_{\bf x,r})\}_{{\bf x}\in
\lambda},
\end{equation}
where the label $a$ stands for the $\bf a$ direction.
More generally, we can consider exchanges of magnitude $\eta \O^{-d}$ with probability 
amplitud $f(\eta)$, the latter being an even function of $\eta$. Needless to say that 
$\phi_{\bf r}$ no longer equals 0,1, and when $\O$ is large enough the value of $\phi_{\bf r}$ 
becomes practically continuous, so that we have
\begin{equation}
C^{\eta{\bf r}a} =\{\phi({\bf x}) + \eta \O^{-d} \nabla_{{\bf x}_a}\d({\bf x- r})
\}_{{\bf x}\in T^d} \quad ,
\end{equation}
where $T^d$ is a $d$-dimensional torus.
A suited Hamiltonian analogue to the internal energy (\ref{ham}) adopts the usual Landau-Ginzburg form
\begin{equation}
H(C) = \O^d \int d^d x\left[ {1 \over 2} (\nabla \phi)^2 + {\t \over 2} 
\phi^2 +{g \over 4!} \phi^4 \right].
\label{lg}
\end{equation}

The occurrence of the factor $\O^d$ in $H(C)$ deserves some comments. It will play
a decisive role when we discuss the influence of the transition rates 
in the dynamics. For
the time being we shall just justify, albeit heuristically, its apperance. 
Let $\xi$ be the system
correlation length. Then, if we choose $1\ll\O < \xi$, the coarse-grained
density will be aproximately constant over distances less than or equal to $\O$.
After carrying partial summations over disjointed regions in $\lambda$ of linear
size $\O$, the factor $\O^d$ emerges and one can then take the continuum limit
by simply making $\O$ large enough.

Lastly, we associate a time dependent statistical weight with each configuration,
$P_t(C)$, which evolves in time accordingly to the following master equation
\begin{eqnarray}
\label{me3}
\partial_t P_t(C)&= &\sum_a \int_R d \eta f(\eta) \int d{\bf r}
 \\ \nonumber
&&\qquad \times \Big\{ W[C^{\eta {\bf r} a} \to C ] P_t(C^{\eta{\bf r}a})-
W[C \to C^{\eta{\bf r}a}] P_t(C) \Big\}.
\end{eqnarray}
The exchange rates $W[C \to C']$ have to be prescribed yet.
We choose them such that 
\begin{equation}
\label{rat}
W[C \to C^{\eta {\bf r}a}]=
D\left(H(C^{\eta {\bf r}a})-H(C)+H_E(C \to C^{\eta {\bf r}a})
\right)\quad ,
\end{equation}
with
\begin{equation}
\label{he}
H_E(C \to C^{\eta {\bf r}a})=\eta {\bf a} \cdot {\bf E}(1-\phi({\bf r})^2)+
O(\e) \quad ,
\label{HE}
\end{equation}
where $\e\equiv\O^{-d}$.  
This means that the transition rates depend on the energy difference
between configurations plus a term whose dominant part in $\e$
is the natural choice to mirror
the effects of the drive as far as it accounts for the local increment of
energy due to the driving field. Correcting terms of higher order in $\e$
will be fixed later.
Again, as in the DLG case described above, in absence of the drive the system
stationary state is the equilibrium one characterized by the $\phi^4$ Hamiltonian
(\ref{lg}).

\subsection{From Fokker-Planck to Langevin}
We are part of the way towards deriving a Langevin equation for the DLG.
The next step we take toward this end is to get a Fokker-Planck equation
by expanding the Master equation in $\e$ up to $\e^2$ order.
To keep matters simple, we will consider E=0 for the moment.
We will recover the $E\not= 0$ case later on.

We avail ourselves of the results collected in appendix A
to expand $H(C^{\eta{\bf r}a})$ and $P_t(C^{\eta{\bf r}a})$ around $C$.
The Master equation transforms into
\begin{eqnarray}
\partial_t P_t(C)=\sum_a \int  d{\bf r} d\eta f(\eta) && \Big\{
\big( D(-\triangle H)-D(\triangle H) \big) P_t(C) \\ \nonumber
&& + \sum_{n=1}^{\infty} {(-\eta \e)^n \over n!} 
\left(\nabla_{{\bf r}_a} {\d \over \d \phi({\bf r})}
\right)^n P_t(C) D(-\triangle H) \Big\},
\label{me1}
\end{eqnarray}
where
\begin{equation}
\triangle H
=\eta \l_a-{\eta^2 \e \over 2} \left( \nabla_{{\bf r}_a} 
{\d \over \d \phi} \right) \l_a + O(\e^2) \quad ,
\end{equation}
with 
\begin{equation}
\l_a=- \nabla_{{\bf r}_a} {\d H \over \d \phi({\bf r})}
\end{equation}

We would like to stress that $\l_a$ is of order one rather than order $\e$.
This is due to the factor $\O^d$ in (\ref{lg}) which 
ensues that our expansion of $D$ is not around zero. This assurance is of the most
importance because it 
bestows dependence upon
the dynamics on equation (\ref{me1}). 
Otherwise we would have found a very different
story: our expansion would
have resulted in a simple Model B \cite{hyh} where any dependence
 on the dynamics
would have vanished. 
Recalling again the results of Appendix A and noticing that much
simplification is obtained because of the integration of odd terms in $\eta$,
one is led to
\begin{equation}
\partial_t P_t(C)= \sum_a  \int d{\bf r} \left(
\nabla_{{\bf r}_a} {\d \over \phi({\bf r})} \right) 
\Big\{\e h(\l_a) P_t(C)+{\e^2 \over 2} e(\l_a) \left( \nabla_{{\bf r}_a}
 {\d \over \phi({\bf r})} \right)P_t(C)
\Big\},
\label{me2}
\end{equation}
where
\begin{eqnarray}
&& h(\l_a)= \int_R \ d\h \ f(\h)\  \h \ D (\h
\l_a), \nonumber \\
&& e(\l_a)= \int_R \ d\h \ f(\h)\  \h^2 \ D (\h
\l_a).
\label{eyh}
\end{eqnarray}
Turning our focus to the $E \not=0$ case, let us choose $
H_E$ in (\ref{HE}) as

\begin{equation}
H_E(C \to C^{\eta {\bf r}a})= \eta \l_a^{(E)}
+\sum_{l=1}^{\infty}{{\eta^{l+1}(-\e)^l}\over{(l+1)!}}
\left( 
\nabla_{{\bf r}_a}
{\d \over \d \phi}
\right)^l \l_a^{(E)}
\end{equation}
with $\l_a^{(E)}= {\bf a} \cdot {\bf E}(1-\phi({\bf r})^2)$.
The point of these manouevres is that the Kramers Moyal expansion is now
trivial. That is, 
one should only substitute in (\ref{me2}) and (\ref{eyh})
$\l_a$ by $\L_a\equiv\l_a +\l_a^{(E)}$.
So, this election allows us to write
\begin{eqnarray}
\partial_t P_t(C)&=& \sum_a  \int d{\bf r} \left(
\nabla_{{\bf r}_a} {\d \over \d \phi({\bf r})} \right)
\Big\{\e h(\L_a) P_t(C) \\
&&\qquad \qquad \qquad \qquad \qquad +{\e^2 \over 2} e(\L_a) \left( \nabla_{{\bf r}_a}
{\d \over \d \phi({\bf r})} \right)P_t(C)
\Big\}.
\nonumber
\end{eqnarray}

We have just derived a Fokker-Planck for the DLG. Now, 
we proceed to find out its stochastically equivalent Langevin equation.
We shall invoke the main result of Appendix B in virtue of which the Fokker-Planck
equation
\begin{eqnarray}
\partial_t P_t(C)=&& \sum_a  \int d{\bf r} \left(
\nabla_{{\bf r}_a} {\d \over \d \phi({\bf r})} \right) \times \nonumber \\
&& \Big\{ f_a(\phi;{\bf r}) P_t(C)+{ 1 \over 2}  \left( \nabla_{{\bf r}_a}
 {\d \over \d \phi({\bf r})} \right)g_a(\phi;{\bf r})^2 P_t(C)
\Big\},
\end{eqnarray}
is equivalent to the Langevin equation
\begin{equation}
\partial_t \phi({\bf r},t)= \sum_{a=1}^d \nabla_{{\bf r}_a} \Big[
f_a(\phi;{\bf r})+g_a(\phi;{\bf r}) \z_a({\bf r},t) \Big],
\end{equation}
using the Ito prescription and with $\z_a({\bf r},t)$ being a gaussian
 white noise, \ie
$\langle \z_a({\bf r},t) \rangle =0$ and $\langle \z_a({\bf r},t) \z_{a'}
({\bf r}',t')\rangle = \d_{a,a'} \d(t-t') \d({\bf r}-{\bf r}')$.
Then, relating to our case, it is straightforward to get
\begin{equation}
\partial_t \phi({\bf r},t)= \sum_a \nabla_{{\bf r}_a} \left[
h(\L_a)+ e(\L_a)^{1/2} \z_a({\bf r},t) \right] \quad ,
\label{lg1}
\end{equation}
where, time has been rescaled by a factor 
$\e$, and finally $\e$ has been set to 1 since no more perturbative expansions
in $\e$ are going to be considered.

Before we proceed further, several comments are in order.
First, the basic symmetries of the DLG are present in the Langevin equation
 (\ref{lg1}):
it is invariant under the simultaneous change $E \to -E$ and $\phi \to -\phi$,
and it is also invariant under translations in space and time.
But the central hallmark is that it depends strongly upon the dynamics. We
believe this is a real step forward, if only because it goes beyond phenomenological
approaches. Gratifyingly, we shall see shortly how to exploit this new state of affairs.

\subsection{Power Counting}
We focus on the critical region where large fluctuations on all length scales
dominate. Further simplification in (\ref{lg1}) is possible in this
regimen by dropping the irrelevants terms in the renormalization group sense. 
Following the standard field theoretic methods let us introduce an 
external momentum scale $\m$ and make the following anisotropic scale transformations:
$t\to \m^{-z} t, r_\pe \to \m^{-1} r_\pe, r_\pa \to \m^{-\s} r_\pa$, and
$\phi \to \m^\d \phi$, where $\pa$ stands for the direction
parallel to the driving field $\bf E$, and $\pe$ for those perpendicular to it.
As usual, the noise scales as $\z_a \to \mu^{(z+d+\s-1)/2} \z_a$.
Next, we expand the Langevin equation in powers of $\m$ around $\m=0$, keeping
only the leading terms.
The time scale, the transverse noise, and the transverse spatial interaction
are forced to remain invariant under the transformation. With this understood
the values of $z$ and $\d$ can be determined. One gets $z=4$ and $\d=(\s+d-3)/2$.
Different scenarios are now possible depending on the value of $\s$.
Demanding that the coefficients of $\nabla_\pe^4$ and $\nabla_\pa^2$ scale in the 
same way, as in the standard analysis of the DLG \cite{jys},
would lead to the choice $\s=2$. 
We show the most representative terms
\begin{eqnarray}
\partial_t \phi &=&-{e(0) \over 2} \bigg[ \sum_\pe \nabla_\pe^4 \phi+\m^{2\s-2}
\D_\pe \D_\pa \phi-\m^{-2} \t \D_\pe \phi -\m^{\s+d-5} {g \over 6} \D_\pe \phi^3
\bigg]\nonumber \\
&&+ h'(E) \bigg[\m^{2\s+2} \D_\pe \D_\pa \phi+ \m^{4\s-4} \D_\pa^2\phi-\m^{2\s-4} \t 
\D_\pa \phi -\m^{3\s+d-7} \D_\pa \phi^3  \nonumber \\
&& \qquad \qquad \qquad-\m^{(3\s+d-11)/2} E \nabla_\pa \phi^2 \bigg]  \nonumber \\
&&+e(0)^{1/2} \sum_\pe \nabla_\pe \z_\pe +\m^{\s-1} e(E)^{1/2} \nabla_\pa \z_\pa,
\label{pcl}
\end{eqnarray}
which after setting $\s=2$ and taking the limit $\m \to 0$, assuming $d > 3$, gives
\begin{eqnarray}
\label{lce}
\partial_t\phi({\bf r})=
&&{e(0)\over{2}}
\biggl[
-\sum_{\pe}\nabla_{\pe}^4\phi+\t\D_{\pe}\phi
+{g\over6}\D_{\pe}\phi^3
\label{lcn}
\biggr] \\ &&-\t 
\ h'(E)
\ \nabla_{\pa}^2 \phi
-E \ h'(E)
\ \nabla_{\pa} \phi^2   
+ \sqrt{e(0)}
\ \sum_{\pe} \nabla_{\pe} \z_\pe({\bf r}).
\nonumber
\end{eqnarray}
$h'$ is the first derivative of the function $h(\L_a)$, closely related
to the first derivative of the transition rate $D$. 

Let us take a glance at the structure of equation (\ref{lcn}).
It can be easily checked that all we have is the Langevin
equation postulated by Leung and Cardy for the DLG and which is often known as
{\em driven diffusive system}. But in stark contrast to \cite{lyc}, equation
(\ref{lcn}) displays the precise form in which the microscopic field enters
the mesoscopic picture of the DLG. More precisely, the two different critical
temperatures introduced by Leung and Cardy for longitudinal and transverse
ordering are indentified here as $\t h'(E)$ and $\t$ respectively, while the
mesoscopic version of the field $\bf E$ finds its counterpart in $E h'(E)$. 
We summarize very briefly now the results of references \cite{lyc,jys}: the upper
critical dimension is shifted from the equilibrium
value $d_c=4$ to $d_c=5$. The reason for this change 
is claimed to lay in the strong anisotropic scaling \cite{syz} ($\s=2$) or, in
other words, the necessity to introduce two length scales associated with
the directions parallel and transverse to the field. An $\e$-expansion can 
be therefore performed in $\e=5-d$ dimensions ---care must be taken of the 
quartic coupling $g$ which is a dangerous irrelevant operator--- leading to critical 
exponents that take their mean-field value (except the anisotropy exponent $\D$ \cite{syz}).
This result is exact, \ie higher order terms in the $\e$-expansion
vanish. This is borne out by a Galilean invariance 
symmetry present in the theory. In particular, 
the order parameter critical exponent $\b$ equals 1/2. This figure does not tally
with $\b \approx 1/3$ which is obtained from Monte Carlo simulations \cite{vym}.
Most of them, with few exceptions, use essentially infinite $E$, \ie no
jumps against the drive are allowed. It has been argued \cite{LW,wan} that an
anisotropic
finite-size scaling analysis of computer simulations leads to critical 
exponents entirely consistent with field theoretic predictions. 
To date, the question remains murky so far an analysis of the data 
in \cite{LW,wan} was exhibited in \cite{MAGA} showing numerical evidence 
for $\b \approx 1/3$.
In the same vein, doubtful data collapse is obtained by all authors for the
two point correlation
function using anisotropic systems, specially at large distances \cite{allz}.
Moreover,  triangle domains pointing opposite to the field are found
by numerically solving (\ref{lcn}) while they are also found in microscopic
simulations, but pointing along the field \cite{allz}!
Several models beyond
the DLG have been proposed, practical motivations aside, to gain
a wider perspective of these problems
\cite{agma,mya} (see also later in this article). Despite
these studies have led to new questions,
some insight into the old ones have been achieved. That is, 
the very existence of a peculiar linear interface has emerged
as a relevant  point in the understanding of the DLG rather than the
presence of a driving field. Also the existence of two different 
length scales is questionable \cite{mya}.

It would be a mistake to dismiss all these puzzles which cast serious 
doubts on the validity of the existing field theory of the DLG.
Originally, it was assumed that an infinite
microscopic electric field implied a finite non-zero coarse grained
driving field \cite{lyc}. In equation (\ref{lce}) we show
explicitely that this is not the case. In fact, it happens there
that when $E=\infty$ the driving term disappears.
We believe that this is the reason of the mismatch between 
the simulational results and the
analysis of the Langevin equation (\ref{lce}).
The latter is not suited for comparison with computer simulations
because these are always performed with infinite drive and in equation
(\ref{lce}) a finite non-zero driving term is present.
Let us take the calculation a stage further by setting
$E$ to infinity in (\ref{lce}). Then, all the terms
depending on the electric field $E$ become identically zero. 
This fact can be easily checked, irrespective of the equation
considered, (\ref{lg1}) before 
power counting or ({\ref{lcn}) after rescaling.
So, (\ref{lcn}) simplifies to
\begin{equation}
\label{Lcn3}
\partial_t \phi({\bf r})=
{1\over{2}}
e(0)\biggl[
-\D_{\pe}^2\phi+\t \D_{\pe}\phi
+{{g}\over{6}} \D_{\pe}\phi^3
\biggr] \nonumber \\ +
\sqrt{e(0)} \
\sum_{\pe}\nabla_{{\pe}}\z_{\pe}({\bf r},t).
\end{equation}

The latter equation constitutes a simple model B \cite{hyh} in the 
transverse directions and no structure in the parallel one.
Thus, the scaling $\s=2$ for $E=\infty$ 
leads to a trivial behaviour. Furthermore, one realizes that imposing
$\nabla_\pe^4$ and $\nabla_\pa^2$ to scale in the same way is
meaningless because there is no parallel gradient term appearing
in the Langevin equation. Nevertheless, we have the freedom to choose $\s$ to 
look for different critical theories. A natural choice is
$\s=1$. With this election, an equation analogue to (\ref{pcl}) can
be written down, however much more involved:
\begin{eqnarray}
\partial_t \phi &=&
-{e(0) \over 2} \bigg[ \sum_\pe \nabla_\pe^4 \phi+
\D_\pe \D_\pa \phi-\m^{-2} \t \D_\pe \phi -\m^{d-4} {g \over 6} \D_\pe \phi^3
\bigg]\nonumber \\
&&h'(E) \bigg[\D_\pe \D_\pa \phi+ \D_\pa^2\phi-\m^{-2} \t
\D_\pa \phi -\m^{d-4} {g \over 6} \D_\pa \phi^3  
-\m^{(d-8)/2} E \nabla_\pa \phi^2 \bigg]  \nonumber \\
&& + h''(E) \bigg[ \mu^{(d-4)/2} \t^2 \nabla_\pa (\nabla_\pa \phi)^2
+\mu^{d-4} {2 \over 3} \t E \D_\pa \phi^3+\mu^{3(d-4)/2} E^2 \nabla_\pa \phi^4 \bigg]
\nonumber \\
&& -\mu^{(5d-16)/2} E^3{h'''(E)\over 6}\nabla_\pa \phi^6 +e(0)^{1/2} \sum_\pe 
\nabla_\pe \z_\pe +e(E)^{1/2} \nabla_\pa \z_\pa.
\label{s1}
\end{eqnarray}
In particular, for $E=0$ one recovers the equilibrium theory (model B). For finite 
$E$, the critical dimension is $d_c=8$ but some of the irrelevant terms 
might be taken into account below the critical temperature. When $E=\infty$
we get a much more simple equation:
\begin{eqnarray}
\partial_t \phi = && {{1}\over{2}}
e(0)
\biggl[
-\D_{\pe}\D_{\pa}\phi-\D_{\pe}^2\phi+\t \D_{\pe}\phi
+ {{g}\over{6}} \D_{\pe}\phi^3
\biggr] \nonumber \\
&& +\sqrt{e(0)} \
\sum_{\pe}\nabla_\pe\z_\pe({\bf r},t)
+{\sqrt{e(0) \over 2}}
\ \nabla_\pa \z_\pa({\bf r},t).
\label{new}
\end{eqnarray}
This equation is the central result of this work. We now proceed to discuss
its physical implications. To begin with, equation (\ref{new}) is
structureless in the parallel direction. So, it corresponds to 
a lattice gas in which particles are exchanged at random in the 
direction of the field while jumps in the transverse directions are 
subject to energetics. 
The most obvious distinction between the finite $E$ case and
the infinite one 
lies in the steady-state current term.
This does not appear in (\ref{new}). 
Of course, such a current exists but it has no bearing on
critical properties. Thus, equation (\ref{new}) does not gather it.

It is quite surprising to find that, for infinite 
driving field, the Langevin equation changes dramatically when compared to
the one that characterizes the finite $E$ case. This would have been 
difficult to work out only on symmetry grounds. 
Also remarkable is the lack of Galilean invariance \cite{jys} 
and the emergence of a single correlation length. 
Anticipating renormalization (the computation of the critical 
exponents will be presented elsewhere), we should remark that
the upper critical dimension is now $d_c=4$ and it yields a 
distinct universality class from that obtained for finite 
fields. Therefore, a value of $\b$ different from $1/2$
is expected.
Finally, we wish to remark that
equation (\ref{new}) is renormalizable when $E= \infty$, while 
for finite values of the field it can be seen that it is not.
A strong crossover from our theory to the finite $E$ case 
occurs for very long impressed fields which obscures the interpretation of 
Monte Carlo data.

\section{Related Models}
\subsection{Two-Temperature model}
In this subsection we consider a model closely related to 
the DLG. The {\em two-temperature Ising lattice gas} \cite{glms} consists of
a set of particles endowed with an Ising Hamiltonian and lying 
on a hypercubical $d$-dimensional lattice.
In contrast to the usual kinetic lattice gas, particle-hole exchanges 
are controlled by rates with different temperatures according to the following 
rules: if the vector $\bf a$ pointing from the particle to the hole 
lies in an 
$n$-dimensional subspace, the ``parallel'' subspace to say, then a transition
rate of the form $D(\triangle H/T_\pa)$ is at play. On the other hand if
$\bf a$ belongs to the ``transverse'' space then the exchanges are coupled to
a bath with temperature $T_\pe$. 
Although this model could be seen somewhat afield from the DLG ---there is
no driving field, for example--- Monte Carlo data for both the DLG and the 
two-temperature
model for $T_\pa=\infty$ look quite similar \cite{cglv}. For instance, 
the value for the order parameter critical exponent in the two-temperature
model is $\b \approx 1/3$, thus sharing the same value
as the DLG with $E=\infty$.
To illustrate this connection further, let us consider our continuum version of the DLG
with infinite drift. As we saw in the previous section this 
corresponds to completely random particle hops in
the field direction. Then we note that the same picture arises if we choose 
$T_\pa= \infty$. So, we should expect 
that the two models lie in the same universality class. 

We now proceed to build a continuum theory for
the two-temperature model.  
Let us start with the master equation 
in continuous space (\ref{me3}). 
Now $W[C \to C']=D\big(H(C')-H(C)\big)$ where
$H(C)$ is the usual $\phi$-four
hamiltonian (\ref{lg}) but it depends on $\bf a$ that the coefficient 
$\tau$ of $\phi^2/2$ will adopt different values.
That is, $\tau=\t_\pe$ if $\bf a$ lies in the transverse subspace
and $\tau=\t_\pa$ otherwise. 
Being initially interested in the $\t_\pa \to \infty$ case, we shall 
assume that particle-hole
exchanges in the parallel subspace are subject to 
\begin{equation}
H=\O^d {\t_\pa \over 2}\int d{\bf x} \phi^2({\bf x}).
\label{tth}
\end{equation}
So, it is expedient to separate (\ref{me3})
into two parts, the first being a sum over the transverse
subspace and the second over the parallel one.
Next, we follow the same steps as in Section 2.2, \ie an expansion
of $\triangle H$ and $P_t(C^{\eta{\bf r}a})$ around $C$.
The calculation is straigtforward, for the 
(simbollicaly) $a \in \pe$ case leads to the same result as in Section
2.2 and when $a \in \pa$ one simply gets $\triangle H= -\eta \t_\pa
\nabla_{{\bf r}_a} \phi$. Then the following Fokker-Planck equation 
can be easily computed
\begin{eqnarray}
\partial_t P_t(C)&=& \sum_{a \in \pe}  \int d{\bf r} \left(
\nabla_{{\bf r}_a} {\d \over \d \phi({\bf r})} \right)
\bigg\{\e h(\l_a)P_t(C)\nonumber \\
&& \qquad \qquad \qquad \qquad \qquad+{\e^2 \over 2} e(\l_a) \left( \nabla_{{\bf r}_a}
{\d \over \d \phi({\bf r})} \right)P_t(C)
\bigg\} \nonumber \\
& +&\sum _{a \in \pa} \int {\e^2 \over 2} e(\t_\pa \nabla_{{\bf r}_a}
\phi) \Bigg( \nabla_{{\bf r}_a} {\d \over \d \phi({\bf r})} \Bigg)^2 P_t(C),
\end{eqnarray}
and in the limiting case $\t_\pa \to \infty$ its equivalent Langevin equation reads
\begin{equation}
\partial_t \phi({\bf r},t)= \sum_{a\in \pe} \nabla_{{\bf r}_a} \left[
h(\l_a)+ e(\l_a)^{1/2} \z_a({\bf r},t) \right] +{\sqrt{e(0)} \over 2}
\sum_{a \in \pa}\nabla_{{\bf r}_a} \z_a.
\end{equation}
Finally, we restrict ourselves to a 1-dimensional parallel subspace. A naive 
dimensional analysis yields, after dropping irrelevant terms and
assuming  $\s=1$,
the same equation we found for the DLG at criticallity with $E=\infty$.
Therefore we are
supplying a firm grasp to support our claimings.
That is, the DLG with $E=\infty$ and the two-temperature model
with $T_\pa =\infty$ are members of the same universality class.

No adiditional effort is necessary to study the finite $T_\pa$ case.
We again seperate (\ref{me3}) into two parts, but now we take 
the full expression for the Hamiltonian (\ref{lg}) rather than 
(\ref{tth}). A derivation running along the same lines of the previous
subsection is possible. We only remind that the scaling $z=4, \s=1$ and 
$\d=(d-2)/2$ is used in the power counting procedure which entails
\begin{equation}
\partial_t \phi = {e(0) \over 2} \Big[ \D^2 \phi +(\t_\pe \D_\pe + \t_\pa \D_\pa) \phi
+{g \over 3!} \D \phi^3 \Big] +\sqrt{e(0)} \sum_a \nabla_{{\bf r}_a} \z_a({\bf r},t).
\label{ttl}
\end{equation}
Despite of the fact that anisotropy is only present as far as the masses are
concerned, we should remark that more general anisotropies 
could be expected. 
Then, the collective behaviour of the 
two-temperature model with finite $T_\pa$ could be more adequately predicted
by means of an extension of (\ref{ttl}) to an
equation with full anisotropy in the coefficients.

\subsection{Randomly Driven Lattice Gases}

Here we take up an extension of the DLG,
the {\em randomly driven lattice gas} (RDLG hereafter) \cite{ran}.
Let us consider a DLG in which the  
driving field fluctuates accordingly to an
even distribution $p[E({\bf x},t)]$ which is $\d$-correlated
in space and time. The RDLG is easier to realize in the
laboratory than the DLG and one can also benefit from
a higher analytical simplicity because a random drive
induces no steady-state current and the particle-hole
symmetry is preserved. 
Like the DLG, it exhibits a second order phase transition 
at half filling
from a disordered 
state to striplike order.
There is hardly any difference between typical
ordered configurations associated with the DLG with an infinite drive 
imposed, the two-temperature model and
the randomly driven model. Now, to
provide a better comparison 
we construct a continuum equation for the RDLG.
Our starting point is again equation (\ref{me3})
with the prescriptions (\ref{rat}) and (\ref{he}).
We should then average over a random $E$,
but we can defer such an average to a later stage.
First, we repeat the steps we performed to arrive at
a Langevin equation for the DLG.
All results carry over without change until we face
to averges over $E$ in (\ref{s1}) of the type
$E^m h^{(n)}(E)$ and $\sqrt{e(E)}$.
As a consequence of a symmetric $p[E({\bf x},t)]$ and the integrations 
over $\eta$, the averages with $m+n$ being an even integer 
vanish whilst the remainder terms will yield the following finite values:
\begin{eqnarray}
\gamma_1 &\equiv& \int dE \ p[E] \ h'(E), \nonumber \\
\gamma_2 &\equiv& \int dE \ p[E] \ E \ h''(E), \nonumber \\ 
\gamma_3 &\equiv& \int dE \ p[E] \ \sqrt{e(E)}.
\end{eqnarray}
The Langevin equation can then be written down in the form
\begin{eqnarray}
\partial_t \phi = -{e(0) \over 2}\sum_{\pe} \nabla_\pe^4\phi +\gamma_1 \D_\pa^2\phi +(\gamma_1-{e(0) \over 2})
\D_\pe \D_\pa \phi +{e(0) \over 2} \t \D_\pe \phi -\t \gamma_1  \D_\pa \phi \nonumber \\
+{e(0) \over 2} {g \over 3!} \D_\pe \phi^3+{2 \over 3} \t (\gamma_2- \gamma_1) \D_\pa \phi^3 +\sqrt{e(0)}
\sum_{\pe} \nabla_\pe \z_\pe +\gamma_3 \nabla_\pa \z_\pa.
\end{eqnarray}
We have arrived at an entirely anisotropic equation, \ie all the gradient operators have been split into
components parallel and transverse to $E$.
In accordance with our earlier discussion of the two-temperature model, we conclude that the RDLG 
and the two-temperature model with finite $T_\pa$ share the same critical behaviour.
But we caution that, as in the DLG, in the RDLG two cases have to be distinguished. We can think of
the simplest distribution $p[E]$, namely the bimodal ${1 \over 2} [\d(E+E_o)+\d(E-E_0)]$, and then take the
limit $E_o \to \infty$. Not surprisingly, in this limit $\gamma_i=0, i=1,2,3$, and
the Langevin equation (\ref{new}) associated with the infinitely driven DLG emerges again. We therefore can conclude
that the generical critical properties of the two-temperature 
model with $T_\pa =\infty$, the DLG with $E$ set to infinity and
the RDLG driven with an effectively infinite field are indistinguishable.

\subsection{Layered Driven Lattices Gases}
We turn our attention to a generalization of the DLG \cite{tllg}.
Let us consider a pair
of identical square lattices placed {\em back to back}. Each plane is a copy
of a two dimensional DLG. No
inter-layer coupling is allowed, but particles can hop from one plane to their
``nearest neighbour'' site in the other one. This process is controlled
by in-plane energetics alone and it is not affected by the drive.
The overall particle density is fixed at 1/2. Concerning the nature of the 
phase transition, Monte Carlo data ($E=\infty$) have revealed that intriguinly 
two transitions appear \cite{tllg}. 
Here, we just give the gist of references \cite{MAGA,agma} where 
the phase diagram in the $(E,T)-$plane
has been mapped out employing Monte Carlo simulations and
dynamic mean-field theory.
As $T$ is lowered the system first orders from a homogeneous state into a state with strips in
both layers. This transition is observed to be characterized by the 
same critical indexes as the DLG. Decreasing $T$ further, we reach the second transition
where homogeneous 
layers with different
density appear.
This transition belongs to the Ising universality class for any
$E < E_c \approx 2$, while values of $E$ beyond the threshold field 
$E_c$ lead to
a first order phase transition \cite{MAGA} .

Although physical motivations for this model come from various directions \cite{MD},
we shall focus on the theoretical side. As far as comparison with simulations is
intended, we shall provide a mesoscopic picture to place into a coherent 
analytical context these two phase transitions.
Next, we construct a continuum mesoscopic theory in the spirit of
Section 2. We first discuss the equilibrium case $E=0$. 
Jumps in this model can be naturally divided into two types:
{\em in}-layer jumps and {\em inter}-layer jumps. So, the 
following notation will prove to be convenient: we shall 
refer to $\phi_1({\bf r})$ and $\phi_2({\bf r})$ as the coarse
grained density field in plane one and two respectively, and
an arbitrary global configuration will be termed 
$C\equiv \{\phi_1({\bf r}),\phi_2({\bf r})\}$. 
We shall denote $C_i^{\eta{\bf r}a}$ the configuration after an
exchange of density $\e \eta$ is performed in the $\bf a$ direction
with an infinitesimal neighbour of $\bf r$ in plane $i$.
Exchanges between planes will lead to configurations
named $C^{\eta{\bf r}}$. More specifically,
\begin{eqnarray}
C_1^{\eta{\bf r}a}&=&\Big\{\phi_1({\bf x}) + \e \eta \nabla_{{\bf x}_a} \d ({\bf x}-
{\bf r}),\phi_2({\bf x})\Big\}, \nonumber \\
\quad C_2^{\eta{\bf r}a}&=&\Big\{ \phi_1({\bf x}),\phi_2({\bf x}) + \e \eta \nabla_{{\bf x}_a} \d ({\bf x}-
{\bf r})\Big\}, \nonumber \\
C^{\eta {\bf r}} &=&\Big\{ \phi_1({\bf x}) +\e \eta \d ( {\bf x}-{\bf r}),
\phi_2 ({\bf x}) -\e \eta \d ({\bf x} -{\bf r}) \Big\}.
\end{eqnarray}
With this understood, then the following master equation can be written down,
\begin{eqnarray}
\partial_t P_t(C)= \sum_a \int d{\bf r} \ d \eta  f(\eta) 
\Big\{ W[C_1^{\eta{\bf r} a} \to C ] P_t(C_1^{\eta{\bf r}a})-
W[C \to C_1^{\eta{\bf r}a}] P_t(C) \nonumber \\
+ W[C_2^{\eta{\bf r}a} \to C ] P_t(C_2^{\eta{\bf r}a})-
W[C \to C_2^{\eta{\bf r}a}] P_t(C)\Big\} \nonumber \\ 
+ \int d{\bf r} \ d\eta \ f(\eta) \ \Big\{ W[C^{\eta, {\bf r}} \to C ] 
P_t(C^{\eta,{\bf r}})-
W[C \to C^{\eta, {\bf r}}] P_t(C) \Big\}. \nonumber \\
\end{eqnarray}
$W[C \to C']$ has its usual meaning, \ie $W[ C \to C']=D\Big( H(C')-H(C) \Big)$
and $H(C)$ is again the Hamiltonian (\ref{lg}).
The calculus towards a Fokker-Planck equation can be carried out as we
did in Section 2, the only difference being the inter-layer current term.
One can easily get 
\begin{eqnarray}
\partial_t P(C)= \sum_{i=1}^2 \sum_a  \int d{\bf r} \left(
\nabla_{{\bf r}_a} {\d \over \d \phi_i({\bf r})} \right)
\Big\{\e h(\l_a^{(i)}) P_t(C) \qquad \qquad \qquad \qquad \\
+{\e^2 \over 2} e(\l_a^{(i)}) \left( \nabla_{{\bf r}_a}
{\d \over \d \phi_i({\bf r})} \right)P_t(C) 
\Big\} && \nonumber \\
+\int d{\bf r} \Bigg(\nabla_{12} {\d \over \phi({\bf r})} \Bigg)
\Big\{ \e h(\l_{12}) P_t(C) +{\e^2 \over 2} e(\l_{12}) \Bigg( \nabla_{12} {\d \over \phi ({\bf r})
} \Bigg) P_t(C) \Big\} &&.
\nonumber
\end{eqnarray}
We now explain our notation.
The first term in this equation models two decoupled Ising 
lattice gases with appropiate constraints. As for the second one, it is simply a
``discrete'' version of the former, as long as it has to be with exchanges across the
layers.
Thus, $\l_a^{(i)}$ has the same meaning as in Section 2,
but restricted to plane $i$. The functions $h$ and $e$ are also defined as 
in the previous section.
With $\nabla_{12} {\d \over \d \phi}$ we simply denote
the operator
$$
{\d \over \d \phi_1({\bf r})} -{\d \over \d \phi_2({\bf r})},
$$
and $\l_{12}$ stands for $\left( \nabla_{12} {\d \over \d \phi}\right) H(C)$.
Turning the drive on, one only has to move $\l_a^{(i)}$ into $\L_a^{(i)}$
thereby taking into account the effect of the electric field.
A Langevin equation can then be derived following the same lines 
of Subsection 2.2. One gets
\begin{eqnarray}
\label{planes}
\partial_t \phi_1({\bf r}) &=&-\G \l_{12} -\G e(\l_{12})^{1/2} \z({\bf r},t)
+\sum_a \nabla_{{\bf r}_a} \Big[ h(\L_a^{(1)})+e(\L_a^{(1)})^{1/2} \z_a({\bf r},t)\Big], 
\nonumber \\
\partial_t \phi_2({\bf r}) &=&  \quad \! \G \l_{12} +\G e(\l_{12})^{1/2} \z({\bf r},t)
+\sum_a \nabla_{{\bf r}_a} \Big[ h(\L_a^{(2)})+e(\L_a^{(2)})^{1/2} \z_a({\bf r},t)\Big].
\nonumber \\
\end{eqnarray}
$\z$ and $\z_a$ are gaussian white noises while $\G$ is a hand introduced
transport coefficient that measures the rate at which the system changes
due to the inter-layer exchange mechanism.

Bearing density conservation in mind, we introduce two new fields:
\begin{equation}
m({\bf x}) \equiv (\phi_1 + \phi_2)/2,
\qquad \varphi ({\bf x}) \equiv (\phi_1-\phi_2)/2.
\end{equation}
Equations (\ref{planes}) are easily expressed in terms of the new 
fields $m({\bf x})$ and $\varphi({\bf x})$.
Now, simulation results hint at which field will be treated as an
order parameter. 
We shall take $m({\bf x})$ as the ordering field
for the DLG type transition, the one that occurs
at a higher temperature. Then, in a naive
dimensional analysis there is much freedom to choose
the scale of observation. We perform the following
scale transformation: $t \to \mu^{-z} t, r \to \mu^{-\s} r, \varphi \to \mu^\d 
\varphi$, and $m \to \mu^\gamma m$. In particular, if we fix the 
exponents $z=4, \s=1, \d=d/2$ and
$\g=(d-2)/2$, it can be easily checked
that, after neglecting terms that are irrelevant in the 
renormalization group sense, we are left with 
nothing but equation (\ref{s1}) for the DLG. Hence, the critical properties 
belong to the DLG universality class, a picture consistent 
with simulations.

Turning next to the second transition, 
we consider $\varphi$ as a non-conserved order parameter.
We propose a critical 
theory naively consistent with $z=2, \s=1,\d=(d-2)/2$ and $\g = d/2$.
This scaling leaves us with a couple of equations
that take the form:
\begin{eqnarray}
\partial_t \varphi &=& \bigg(1 -{\t h'(0)\over 2 \G}\bigg)\D_\pe \varphi
 + \bigg(1-{\t h'(E)\over 2\G}\bigg) \D_\pa \varphi -\mu^{-2} \t \varphi -
\mu^{d-4} 
{g \over 6} \varphi^3
\nonumber \\
&&- \mu^{d-2} {g \over 2} \varphi m^2
-\sqrt{e(0)} \z, \nonumber \\
\partial_t m &=& \t \Big(h'(0)\D_\pe+h'(E)\D_\pa\Big) m
-\mu^{d-2} {g \over 2} \Big( h'(0) \D_\pe 
+ h'(E) \D_\pa \Big) (\varphi ^2 m) \nonumber \\
&& -\mu^{(d-2)/2} E h'(E) \nabla_\pa m^2-\mu^{(d-6)/2}E h'(E) \nabla_\pa \varphi^2
\nonumber \\ 
&&  +\sqrt{e(0)} \sum_\pe \nabla_\pe (\z_\pe^{(1)} +\z_\pe^{(2)}) +
\sqrt{e(E)} \nabla_\pa (\z_\pa^{(1)} +\z_\pa^{(2)}).
\label{modc}
\end{eqnarray}
We have dropped all terms that give a negligible contribution
in the limit $\mu \to 0$. 
Due to the electric field, which singles out a lattice axis,
all gradient terms have become anisotropic.
The situation is then very much reminiscent of the driven lattice 
gases with repulsive interactions \cite{lsz}.
We have an electric field $\bf E$ that
has an effect on the phase transition only through an
auxiliary non-ordering field $m({\bf x})$.
The naive dimension of E turns out to be $(d-2)/2$, in contrast to
$(d-8)/2$ (Section 2.3), so it is highly irrelevant compared 
to $g \varphi^3$.
We note that essentially the same set of equations results in \cite{lsz}, 
so our analysis of (\ref{modc}) will follow the same lines of this reference. That is, 
$\bf E$ is naively irrelevant for the Gaussian 
fixed point until $d$ is lower than two. 
We conjecture that $E$ is not relevant to 
the Wilson-Fisher fixed point for $d<4$,
so the unique effect of the drive consists in generating anisotropies. 
It should be noted that the field $m({\bf x})$ does not order, so we do not need to keep track of
it as far as critical behaviour is concerned. 
Thus, the critical properties for the two-layer driven lattice gas 
are given by the Langevin equation for $\varphi({\bf x})$ 
and they fall into the Ising universality class. However, corrections to 
order $O(E^2)$ show that $g$ and $\t$ decrease to an amount that depends on $E$.
Eventually, both of them may vanish simultaneously, a mechanism 
that would be liable for a tricritical point. Then, in qualitative agreement with simulations, 
the transition would be discontinous
for $E$ beyond a critical field value. 
Interestingly enough, it could be worked out the dependence 
of the transition temperature on the dynamics.

\section{Summary and Conclusions}
Here, we take stock of what we have done.
Section 1 outlines the task undertaken in this article: to arrive at a field theoretic description
of the DLG model and to show how this description explains why the DLG,
and three models intimately related to it,
have the various properties that they exhibit.
Inspired by the dynamics at the microscopic level, we were able to 
construct a master equation in continuous space. The goal of the avenue we have pursued 
was to take into account the microscopic details, an effort towards
disentangling the role of dynamics in non-equilibrium critical
behaviour.
An expansion of the continuous master equation was then possible due to the factor $\O^d$, a remnant
of the coarse graining over the underlying lattice. The role of the factor $\O^d$ can only be fully appreciated
noticing that all dependence on the dynamics relies on it.
Then, the full Langevin equation for the DLG was derived.
The richness of equation (\ref{lg1}) was brought to fruition when viewed on different length 
scales, leading to the emergence of distinct critical theories. In particular, we
have recovered the Langevin equation of reference \cite{lyc} after considering an 
anisotropic scale transformation ($\s=2$). This result, interesting though it is, is
unsatisfactory in several respects. Firstly, simulations favour a value
$\b \approx 1/3$ for the order parameter critical exponent, whilst
a mean field value $\b =1/2$ ensues exactly in \cite{lyc} for $2< d\leq 5$. 
Diming further the validity of equation (\ref{lce}) as a continuum description of the DLG, numerical 
work on equation (\ref{lce}) conflicts with the simulation data \cite{allz}.
The situation became more transparent after we resorted to the detailed dependence 
of the coefficients on the microscopic dynamics. That is, the case $E=\infty$, which is
with few exceptions the most studied one, was carried out explicitely the result being
a trivial equation. This is a consequence of the choice $\s=2$, which makes no sense
as it was seen in Section 2. Following well honed arguments \cite{mya} that invoke
a single effective correlation length, we have turned to the choice $\s=1$.
Possibly against intuition, different critical behaviour has been found for
the finite $E$ and $E=\infty$ cases. The upper critical dimension associated 
with the former is $d_c=8$ whereas the latter is characterized by $d_c=4$,
thereby in either case yielding an universality class other than that
obtained in \cite{lyc}.

In a subsequent section we have provided three examples of the applicability
of our methods in the shape of the two-temperature model, the
random driven lattice gas and the driven
bi-layer lattice gas.
On symmetry grounds we have suggested why the DLG and the two-temperature model
with $T_\pa =\infty$ should lay in the same universality class. Our suspicions 
have been confirmed so far a Langevin equation identical to the one associated
with the DLG has resulted in a derivation that runs along the
same lines of Section 2. 
The finite $T_\pa$ case has also been studied. Different critical behaviour
from the $T_\pa=\infty$ case has been found. Remarkably, the RDLG model
with finite averaged external field has resulted in a Langevin equation identical 
to that associated with the two-temperature model with $T_\pa$ finite, whilst
the RDLG with an infinite averaged driving field was proved to belong 
to the same universality class of the infinitely driven DLG. Summing up, we conclude
that the infinitely driven DLG, the two-temperature model with $T_\pa =\infty$
and the RDLG with infinite drive, are described by the same Langevin
equation. On the other hand, the two-temperature model with finite $T_\pa$ and
the RDLG with finite averaged field belong into the same universality class.
 
In the last part of Section 3 we have tackled with the two-layer driven lattice gas.
A viable explanation for the two transitions exhibited in this model has been provided in the frame
of field theory. Again, the election $\s=1$ has proved to suffice our purposes, \ie
the understanding of collective behaviour in these systems.
Clearly much work is still to be done as more detailed 
studies on these subjects would be highly
desirable.

\appendix
\section{Formal developements}
Supose that $F$ is a functional of a function $\phi(x)$.
If $\phi$ changes to $\phi +\d \phi$, then a Taylor series 
expansion can formally be written down
\begin{eqnarray}
F(\phi +\d \phi)= F(\phi)&+ &\int dx_1 {\d F \over \d \phi(x_1)} 
\d \phi(x_1)  \\
&& +{1 \over 2} \int dx_1 dx_2 {\d^2 F \over \d \phi(x_1) \d \phi(x_2)}
\d \phi(x_1) \d \phi(x_2) + \ldots,
\nonumber
\end{eqnarray}
where $\d F / \d \phi$ means the functional derivative of $F(\phi)$ with respect
to $\phi(x)$. Now, we explicitely treat the case we are concerned with, namely
\begin{equation}
\d \phi =\e \nabla_x \d(x-r).
\end{equation}
In such a case it inmediately follows that
\begin{eqnarray}
F(\phi +\d \phi)&=&F(\phi) + \sum_{n=1}^\infty {1\over n!} \int dx_1 \ldots dx_n
{\d^n F(\phi) \over \d\phi(x_1) \ldots \d \phi(x_n)} \nonumber \\
&& \qquad \qquad \times
 \e^n \nabla_{x_1} \d(x_1-r) \ldots \nabla_{x_n} \d(x_n-r) \nonumber \\
&=&F(\phi)+\sum_{n=1}^\infty {\e^n \over n!} \bigg\{ \prod_{k=1}^n \int dx_k \nabla_{x_k}
\d(x_k-r) {\d \over \d \phi(x_k)} \bigg\} F(\phi) \nonumber \\
&=& F(\phi)+\sum_{n=1}^\infty {\e^n \over n!} \bigg(  \nabla_r
 {\d \over \d \phi(r)} \bigg)^n F(\phi).
\end{eqnarray}
The operator $\big( \nabla_r {\d \over  \d \phi}\big)$ satisfies 
\begin{equation}
\bigg( \nabla_r {\d \over \d \phi} \bigg) F(\phi)=
\nabla_r \bigg({\d F(\phi) \over \d \phi(r)} \bigg) - {\d \over \d \phi(r)}
\Big( \nabla_r F(\phi) \Big),
\end{equation}
which can be better proved by putting the last expression in a lattice.
Finally, the usual properties of functional derivatives can be applied,
$v.g.$ 
\begin{equation}
\bigg( \nabla_r {\d \over \d \phi} \bigg) F_1(\phi)F_2(\phi)=
F_1(\phi) \bigg( \nabla_r {\d \over \d \phi(r)} \bigg) F_2(\phi) 
+ F_2(\phi) \bigg( \nabla_r {\d \over \d \phi(r)}
\bigg) F_1(\phi).
\end{equation}
\section{Fokker-Planck and Langevin equations for systems with
conserved order parameter}
A Langevin equation with conserved order parameter has the general form
\begin{equation}
\partial_t \phi(x,t) = \sum_a \nabla_{x_a} \left[ f_a(\phi,x)+
g_a(\phi,x) \z_a(x,t) \right],
\end{equation}
$\z_a$ being a gaussian white noise:  
$\langle \z_a(x,t) z_a'(x',t') \rangle= \d_{a,a'} \d(x-x') \d(t-t')$
; $\langle \z_a(x,t) \rangle =0$.
Let us introduce a time discretization. Then
\begin{equation}
\phi_{n+1}=\phi_n + \epsilon \sum_a \nabla_{x_a} \left[
f_a(\phi_n,x)+g_a(\phi_n,x)\z_{a,n} \right],
\end{equation}
where $\phi_n \to \phi_t$ and $t_n=n \epsilon \to t$ when $\epsilon \to 0$
and $n\to \infty$. A factor $\epsilon$ has been absorbed into the noise and 
we are using the Ito prescription. 

The probability that the system is in the configuration $\phi_{n+1}(r)$ at time
$t_{n+1}$ is given by
\begin{equation}
P_{n+1}(\phi_{n+1})= \Big\langle \int d\phi_n P_n (\phi_n) \ \d (S_n) \Big\rangle_\z,
\end{equation}
\begin{eqnarray}
\d (S_n)\approx \d(\phi_{n+1}-\phi_n)+ \int dx_1 {\d \d(\phi_{n+1}-\phi_n) \over
\d \phi_n} \Big[ -\epsilon \sum_a \nabla_{x_a} [f_a+g_a\z_a]\Big]+\nonumber \\
+{1 \over 2} \int dx_1 dx_2 {\d^2 \d(\phi_{n+1} -\phi_n) \over \d \phi_n(x_1)
\d \phi_n(x_2)} \Big[ -\epsilon \sum_a \nabla_{x_{1,a}} \ldots \Big] \
\Big[ -\epsilon \sum_a \nabla_{x_{2,a}} \ldots \Big] +\ldots
\end{eqnarray}
The calculus now reduces to noise averages. An intermediate step is
\begin{eqnarray}
\epsilon^{-1} (P_{n+n}-P_n)= -\sum_a \int dx_1 {\d \over \d \phi} 
\big[ P_n \nabla_{x_a} f_a
\big]+ \nonumber \\ 
{1 \over 2} \int dx_1 dx_2 {\d^2 \over \d \phi(x_1) \d \phi(x_2)}
\Big[ P_n \sum_a \nabla_{x_{1,a}} \nabla_{x_{2,a}} (g_a^2 \ \d(x_1-x_2))\Big],
\end{eqnarray}
and after a bit of algebra, applying the results of Appendix A, our final result reads
\begin{equation}
\partial_t P_t(\phi)= \sum_a \int dx \bigg( \nabla_{x_a} {\d \over
\d \phi(x)} \bigg)\bigg[P_t f_a+{1 \over 2} \Big(\nabla_{x_a} {\d \over
\d \phi(x)} \Big)(P_t \ g_a^2) \bigg].
\end{equation}


\begin{thebibliography}{9}
\bibitem{grins}
G. Grinstein, J. Jayaprakash and Y. He, {\em Phys. Rev. Lett.\/} {\bf 55}, 2527 (1985).

\bibitem{MD} 
J. Marro and R. Dickman,  {\it Nonequilibrium Phase Transitions 
in Lattice Models
}, Cambridge University Press, (Cambridge, 1998).

\bibitem{KLS} 
S. Katz, J.L. Lebowitz and H. Spohn,
{\em Phys. Rev.\/} {\bf B28}, 1655 (1983); {\em J. Stat. Phys.\/} {\bf 34}, 497 (1984).

\bibitem{vym}
J.L. Vall{\'e}s and J. Marro, {\em J. Stat. Phys.\/} {\bf 49}, 89
(1987).

\bibitem{lyc} K.t. Leung and J. Cardy, {\em J. Stat. Phys.\/} {\bf 44}, 567 (1986).
   
\bibitem{jys}
H.K. Janssen and B. Schmittmann, {\em Z. Phys.\/} B {\bf 64}. 503
(1986).

\bibitem{MAGA} J. Marro, A. Achahbar, P.L. Garrido and J.J. Alonso,
{\em Phys. Rev.} E {\bf 53}, 6038 (1996).

\bibitem{LW} K.-t. Leung, {\em Phys. Rev. Lett.\/} {\bf 66}, 453 (1991); J.S.
Wang, {\em J. Stat. Phys.\/} {\bf 82}, 1409 (1996).

\bibitem{allz}
F.J. Alexander, C.A. Laberge, J.L. Lebowitz and R.K.P. Zia, {\em J. Stat. Phys.\/} 
{\bf 82}, 1133 (1996).

\bibitem{pre}
P. L. Garrido, F. de los Santos and M. A. Mu\~noz, 
{\em Phys. Rev. \/} E {\bf 57}, 752 (1998).

\bibitem{gas}
C.N. Yang and T.D. Lee, {\em Phys. Rev.\/} {\bf 87}, 404 (1952).

\bibitem{syz}
B. Schmittmann and R. K. P. Zia, {\em Statistical Mechanics of
Driven Diffusive Systems\/}, in Phase Transitions and Critical 
Phenomena, eds. C. Domb and J. Lebowitz (Academic, London)

\bibitem{hyh}
P.C. Hohenberg and B.J. Halperin, {\em Rev. Mod. Phys.\/} {\bf 49}, 435 (1977).

\bibitem{wan}
J.S. Wang, {\em J. Stat. Phys.\/} {\bf 82}, 1409 (1996).

\bibitem{agma}
J.J. Alonso, P.L. Garrido, J. Marro, A. Achahbar, {\em J. Phys.\/} A {\bf 28},
4669 (1995).

\bibitem{mya}
J. Marro, A. Achahbar, {\em J. Stat. Phys.\/} {\bf 90}, 817 (1998).

\bibitem{glms}
P.L. Garrido, J.L. Lewowitz, C. Maes and H. Spohn, {\em Phys. Rev.} 
A {\bf 42}, 1954 (1990).

\bibitem{ran}
B. Schmittmann and R.K.P. Zia, {\em Phys. Rev. Lett.\/} {\bf 66}, 357 (1991).

\bibitem{cglv}
Z. Cheng, P.L. Garrido, J.L. Lebowitz and J.L. Vall\'es, {\em
Europhys. Lett.\/} {\bf 14}, 507 (1991).

\bibitem{tllg}
A. Achahbar and J. Marro, {\em J. Stat. Phys.\/} {\bf 78}, 1493 (1995).
\bibitem{hzs}
C.C. Hill, R.K.P. Zia and B. Schmittmann, {\em Phys. Rev. Lett.\/} {\bf 77},
514 (1996).

\bibitem{lsz}
K.-t. Leung, B. Schmittmann and R.K.P. Zia, {\em Phys. Rev. Lett.\/}
{\bf 62}, 1772 (1989).

\end{thebibliography}
\end{document}